----------------------------Latex File--------------------------------------

\documentstyle[twoside,fleqn,espcrc2,epsf]{article}
\newcommand{\as} {\alpha_{s}}
\newcommand{\barb} {\overline b}
\newcommand{\bart} {\overline t}
\newcommand{\barf} {\overline f}
\newcommand{\barq} {\overline q}
\newcommand{\mt} {m_{top}}
\newcommand{\mh} {m_{Higgs}}
\newcommand{\mz} {m_{Z}}
\newcommand{\mb} {m_{b}}
\newcommand{\mc} {m_{c}}

\newcommand{\rb} {r_{b}}
\newcommand{\Z} {Z}
\newcommand{\ord} {\cal O}
\newcommand{\see} {\sin^2\theta_{e}}
\newcommand{\seb} {\sin^2\theta_{b}}
\newcommand{\wz} {\Gamma_{Z}}
\newcommand{\wh} {\Gamma_{h}}
\newcommand{\we} {\Gamma_{e}}

\newcommand{\afb} {A_{FB}}
\newcommand{\AmS}{{\protect\the\textfont2
  A\kern-.1667em\lower.5ex\hbox{M}\kern-.125emS}}

\title{The $\Z$ Line-shape in the Standard Model\thanks{Talk given
at the 1994 Zeuthen Workshop on Elementary Particle Theory}\hfill DFTT/G-94-1}

\author{Giampiero Passarino\address{Dipartimento di Fisica Teorica,
Universit\`a di Torino}\address{INFN, Sezione di Torino}}

\begin{document}

\begin{abstract}
The status of the theoretical uncertainties for LEP~1 observables associated
with the corresponding comparison among different codes is briefly reviewed.
\end{abstract}

\maketitle

The achieved experimental accuracy at LEP~1 requires a detailed comparison
of the theoretical predictions from different codes with the final goal of
an estimate of the theoretical error. This is presently under investigation
by the LEP~1 Precision Calculations Working Group and here a short report
will be given on the status of the $\Z$ parameters.

The comparison will proceed in three different phases by considering the
results of five different codes, MIZA~\cite{miza}, LEPTOP~\cite{leptop},
TOPAZ0~\cite{topaz0}, WOH~\cite{miza} and ZFITTER~\cite{zfitter}. In phase
1 one has taken into account the predictions for the so called
pseudo-observables, namely $\see, \seb, \dots$, while in a second step
the $\Z$ parameters have been compared, i.e. $\wz,\wh,\we, \dots$ or
$R,\rb, \dots$ or $\afb^0(l), \dots$. The final step in this project will
require a comparison at the level of realistic observables, like
$\sigma(e^+e^- \to \barf f)$ and $A_{FB}(e^+e^- \to \barf f)$, including a
realistic set-up and $f = e$. At present the first two phases have been
completed.
As a result of this work we have realized the fact that there
are two complementary objectives. In one case the theoretical uncertainty
for a given observable $O$ is estimated as

\begin{equation}
\Delta O = \frac{1}{2} (\max_{i=Codes} - \min_{i=Codes})\,O_i
\end{equation}

On the other end we can just derive additional informations from each single
code by adopting different options which in turns are related to different
implementations of higher orders radiative corrections.
In the end we propose
to compare bands of predictions instead of lines of predictions, as a function
of $\mt$ or of any other unknown parameter of the standard model. In the
following we would like to present a short list of Options with some words
of comments.

\vspace{.2cm}
\begin{figure}[htbp]
\begin{center} \mbox{
\epsfysize=7.5cm
\epsffile{cge.ps}}
\end{center}
\caption[
Predictions for $\Gamma_e$.
]{\label{fig1}
{
Predictions for $\Gamma_e$.
}  }
\label{fig1}
\end{figure}

The scale of $\alpha$ in the final state QED corrections. This is the
well known factor $1 + \frac{3}{4}Q_f^2\frac{\alpha}{\pi}$. Actually this
should not be an Option at all since the correct scale is
$\alpha(\mz)$~\cite{cos},
i.e. the correction factor is

\begin{equation}
1 + \frac{\alpha(\mz)}{\pi}Q_f^2[\frac{3}{4}
- \frac{1}{4}\frac{\alpha_s(\mz)}{\pi}] + \ord(\alpha^2)
\end{equation}

The scale in the vertex EW corrections. We simply mention the three
possibilities, namely $\alpha(0)$, $\alpha(\mz) \equiv G_F$ or $G_F$ for
the leading (sub-leading) $\mt$ corrections and $\alpha(0)$ otherwise.
The scale in the FTJR~\cite{ftjr} corrections to $\Z \to \barb b$.
At the moment the common choice is $\as(\mt)$, i.e.

\begin{eqnarray}
\Gamma_{Z\barb b} &=& \Gamma_0\,\{ 1 + \frac{\as(\mz)}{\pi} \\
&-& \delta [ 1 + (3 - \pi^2)\,\frac{\as(\mt)}{\pi}]\}
\end{eqnarray}

but one could also isolate the gluon radiation as

\begin{eqnarray}
\Gamma_{Z\barb b} &=& \Gamma_0\{1 + \frac{\as(\mz)}{\pi}\} \\
&\times& \{1 - \delta[1 - \pi^2\,\frac{\as(\mt)}{\pi}]\}
\end{eqnarray}

and the QCD factor $1 + \as(\mz)/\pi$ is not included for
the asymmetry $A_{FB}(b)$ and for $\sin^2\theta_b$.
The aimed accuracy requires that $\mb \neq 0$ also in one loop
diagrams. At the moment what it is used is

\begin{equation}
\Gamma_{Z\barb b} = {{G_F\mz^3}\over {8\sqrt{2}\,\pi}}\,\rho\,
\{ g_V^2 + (1 - 6\,{{\mb^2}\over {\mz^2}})\,g_A^2\}
\end{equation}

However vertex corrections, included in $g_{V,A}$ are usually computed for
$\mb = 0$ and this is not fully consistent even if in the missing contributions
the leading terms $\ord(\mb^2\mt^2)$ are absent. Moreover it is an open
question what to use for $\mb$ in this case, the pole mass $m_b = 4.7\,$
GeV  or the running one ${\overline m}_b(\mz)$?

\vspace{.2cm}
\begin{figure}[htbp]
\begin{center} \mbox{
\epsfysize=7.5cm
\epsffile{cgb.ps}}
\end{center}
\caption[
Predictions for $\Gamma_b$.
]{\label{fig2}
{
Predictions for $\Gamma_b$.
}  }
\label{fig2}
\end{figure}

The physical Higgs contribution to the correction factor $\Delta\rho$
is not ultra-violet finite and only through the $\overline{MS}$ prescription
we have a finite $\Delta\rho_H(\overline{MS})$. After that we are left with
the option of a re-summation of such contribution, which makes the result
slightly scale dependent. Otherwise we can just decide that all bosonic
corrections are expanded to first order.

The scale of $\as$ in the $\ord(\alpha\as)$ corrections to the vector
boson self-energies~\cite{aas}. This question is particularly relevant in
view of
a correct treatment of the $\bart t$ thresholds~\cite{ttbth} where it has
been recently suggested~\cite{sv} that the non-perturbative effects can be
numerically recovered by allowing a relatively small scale in the perturbative
expansion, i.e. $\as(0.154\mt)$.
The singlet QCD contribution which is simple and unambiguous for the
hadronic width but which becomes ambiguous, starting at $\ord(\as^2)$,
for individual $\barq q$ channels. Actually some sort of agreement has been
recently reached on these matter but we want to summarize the roots of
the problem~\cite{qcd}. From a pragmatic point of view there is a hierarchical
description where

\begin{equation}
\Gamma_{\barq q} = \Gamma(\Z \to \barq q(g) + \barq' q')
\end{equation}

for all $q'$ such that $m_{q'} < m_q$. On the other end we could have a
democratic description where the final states $\barq q + \barq' q'$ are
assigned for $\frac{1}{2}$ to $\Gamma_{\barq q}$ and for the other
$\frac{1}{2}$
to $\Gamma_{\barq' q'}$. The two descriptions agree fortunately for the
leading terms. In general one could also decide that such final states should
not be assigned to any specific channels in such a way that

\begin{eqnarray}
\wh &\neq& \sum_q\,\Gamma_{\barq q} = \sum_q\,\Gamma(\Z \to \barq q) \\
&+& \sum_{q,q'}\,\Gamma(\Z \to\barq q \barq' q') + \dots
\end{eqnarray}

In particular there is an $\ord(\as^3)$ contribution to $\Gamma_V^S$ which
cannot be assigned to any specific flavour.

\vspace{.2cm}
\begin{figure}[htbp]
\begin{center} \mbox{
\epsfysize=7.5cm
\epsffile{cds2e100.ps}}
\end{center}
\caption[
Absolute deviations for $\sin^2\theta(e)$ from the average value,
$\mh = 100\,$GeV.
]{\label{fig3}
{
Absolute deviations for $\sin^2\theta(e)$ from the average value,
$\mh = 100\,$GeV.
}  }
\label{fig3}
\end{figure}

Perhaps the most important effect is related to the missing EW
higher orders. Typically let $g_{V,A}$ be the vector or axial-vector
coupling of the $\Z$ to fermions. We write

\begin{equation}
g_i = g^0_i + \frac{\alpha}{\pi}\, g^1_i + \ord(\alpha^2)
\end{equation}

where $i= V,A$ and where $g^0_i$ is including the re-summation of universal
terms. In computing partial widths or deconvoluted asymmetries different
codes adopt different choices, i.e.

\begin{eqnarray}
g_i^2 &=& (g^0_i + \frac{\alpha}{\pi}\, g^1_i)^2  \\
{} &=& (g_i^0)^2 + 2\,\frac{\alpha}{\pi}\,g_i^0g_i^1
\end{eqnarray}

where in the first case we square numerically and in the second one we expand
consistently in perturbation theory. This two Options lead to sizeable effects
if we consider for instance $\afb^0$.

Another Option which should be taken into account is relative to the
factorization vs non-factorization of final state QCD corrections. For a
given channel one can write

\begin{equation}
\Gamma_{\barq q} = {{G_F\mz^3}\over {8\sqrt{2}\,\pi}}\,\rho\,
( g_V^2 + g_A^2)\,( 1 + \delta_{QCD})
\end{equation}

where $g_V,g_A$ include non-universal vertex corrections.
Notice that for $b$-quarks, in order to avoid double-counting, the FTJR term
must not include $1+\as/\pi$. However the complete answer at
$\ord(\alpha\as)$ is not known and strictly speaking the QCD correction
should only multiply the universal terms absorbed into $g_V$ and $g_A$.
Thus one can also adopt a non-factorized width both for $b$ and light
quarks. We mention also that the double scale in the FTJR term, while
certainly gauge-invariant for the leading $\mt$ corrections, remains
questionable for the sub-leading and constant ones.
Indeed, going back to the $\ord(\alpha\as)$ corrections, we recall that
the full result, even including $\mb \neq 0$, is available

\begin{equation}
\alpha\as\,\mt^2 [ 1 + \frac{K}{\mt^2} + \ord(\frac{\mz^2}{\mt^4})]
\end{equation}

while for vertex corrections to $\Z \to \barq q$ the following results
are known: for $q\neq b$ the $\ord(\alpha\as\,const.)$ is missing, for
$q = b$ the leading $\alpha\as\mt^2$ corrections is the well known
FTJR term while all the sub-leading (log and non-log) terms are missing.

The present choice for the comparisons is $1/\alpha(\mz)|_{light} =
128.87 \pm 0.12$~\cite{priv} but unfortunately the full updated analysis,
including the behavior of $\alpha(p^2)$, has not yet been published.

The running of $\mc$ has been included in the numerical work but it
should be noticed that the use of

\begin{eqnarray}
{\overline m}(\mc) &=& \mc\,[ 1 - \frac{4}{3}\,x(\mc)  \\
&+& (\frac{16}{9} - K_c)\,x^2(\mc)]
\end{eqnarray}

where $x = \frac {\as}{\pi}$ gives unreliable results due to the low scale
needed in $\as$. We suggest therefore to use

\begin{equation}
{\overline m}(\mc) = {{\mc}\over {1 + \frac{4}{3}\,x(\mc) + K_c\,x^2(\mc)}}
\end{equation}

where $\mc = 1.5\,$GeV is the pole mass.

In the light of the fact that $\mt \approx 1.9\,\mz$ it looks opportune to
rise the question whether or not the full two-loop standard model predictions
are needed and requested. There are two possible answers, namely the present
experimental accuracy plus the uncertainties connected with QED strongly
support the idea that we don't need a full two-loop calculation.
Actually there are discrepancies among various $\ord(\alpha^2)$ QED
Bhabha generators that should be solved before devoting any attempt
towards the two-loop electroweak effects.
To the contrary of that a certain requirement of internal consistency,
especially illustrated by our choice of options, would suggest that we indeed
need them.

In conclusion we can affirm that by comparing the available semi-analytical
codes which are based, among other things, on different renormalization schemes
(${\overline{MS}}$ or on-shell) or the same code with different Options for
radiative corrections we derive the result that

\begin{equation}
\Delta_{th}O \ll \Delta_{exp}O
\end{equation}

where $O$ is any pseudo-observable or $\Z$ parameter with the inclusion of QCD
corrections. Certainly the remaining differences must be investigated both
for internal consistency and for aesthetical reasons. Hopefully new
calculations will contribute in a near future to make some of the Options
obsolete and to unify the various treatments of radiative corrections. On the
other hand a certain tendency to create a common default, by adopting a common
procedure in front of alternative possibilities, should be taken with the due
caution.

\vspace{.2cm}
\begin{figure}[htbp]
\begin{center} \mbox{
\epsfysize=7.5cm
\epsffile{cds2e1000.ps}}
\end{center}
\caption[
Absolute deviations for $\sin^2\theta(e)$ from the average value,
$\mh = 1\,$TeV.
]{\label{fig4}
{
Absolute deviations for $\sin^2\theta(e)$ from the average value,
$\mh = 1\,$TeV.
}  }
\label{fig4}
\end{figure}

For cross sections and asymmetry with a realistic set-up the work is in
progress among those groups with libraries which are also QED-dressers
(BHM, TOPAZ0 and ZFITTER), but the greatest effort must be spent in order
to reduce the theoretical error. Certainly the first step will be a detailed
comparison for the forward-backward asymmetry $\afb$.

Finally a very small sample of the various comparisons is shown in Figures 1-5.
In order to give an idea of the situation we have decided to plot in Figure 1
the various predictions for $\we$ as a function of $\mt$.
The same information is displayed in Figure 2 for $\Gamma_b$.
For a better understanding of the differences among the four codes we present
in Figure 3-4) the absolute deviation of the four codes for $\sin^2\theta(e)$
from its average.
In Figure 5 we present the theoretical uncertainty for the
ratio $R$ as estimated by TOPAZ0 by switching on and off the various options
that we have discussed above.

\vspace{.2cm}
\begin{figure}[htbp]
\begin{center} \mbox{
\epsfysize=7.5cm
\epsffile{br.ps}}
\end{center}
\caption[
Predictions and uncertainties for $R$ from TOPAZ0.
]{\label{fig5}
{
Predictions and uncertainties for $R$ from TOPAZ0.
}  }
\label{fig5}
\end{figure}

I would like to thank Oreste Nicrosini, Guido Montagna and Fulvio
Piccinini for the continuous collaboration. I am also grateful to
Dima Bardin and Manel Martinez for stimulating discussions. Finally
I would like to thank Tord Riemann for the invitation and the very
pleasant atmosphere at this Conference.


\begin{thebibliography}{9}
\bibitem{miza} G.~Burgers, W.~Hollik and M.~Martinez, program BHM;
W.Hollik, Fortschr. Phys. 38 (1990) 3, 165;
M.Consoli, W.Hollik and F.Jegerlehner: Proceedings of the Workshop on
Z physics at LEP I, CERN Report 89-08 Vol.I,7;
G.Burgers, F.Jegerlehner, B.Kniehl and J.H.K{\"u}hn: the same proceedings,
CERN Report 89-08 Vol.I,55.

\bibitem{leptop} V.~A.~Novikov, L.~B.~Okun, A.~N.~Rozanov and
M.~I.~Vysotsky, CERN preprint CERN-TH.7217/94.

\bibitem{topaz0} G. Montagna, O. Nicrosini, G. Passarino, F. Piccinini
and R. Pittau, Nucl. Phys. B401(1993)3.
G. Montagna, O. Nicrosini, G. Passarino, F. Piccinini and R. Pittau,
Comput. Phys. Commun. 76(1993)328.

\bibitem{zfitter} D.~Bardin et al., program ZFITTER 4.0;
Nucl. Phys.  B351 (1991) 1;
Z. Phys. C44 (1989) 493;
Phys. Lett.  B255 (1991) 290.

\bibitem{cos} A.~L.~Kataev, Phys. Lett. B287()209.

\bibitem{ftjr} J.~Fleischer, O.~V.~Tarasov, F.~Jegerlehner and P.~Raczka,
Phys. Lett. B293 (1992) 437.

\bibitem{aas} B.~A.~Kniehl, Nucl.Phys. B347(1990)86;
A.~Djouadi, Nuovo Cim. 100A(1988)357.

\bibitem{ttbth} V.~S.~Fadin and V.~A.~Khoze, JETP Lett. 46(1987)525;
B.~A.~Kniehl and A.~Sirlin, Phys. Rev. D47(1993)883;
F.~J.~Yndurain, Madrid preprint FTUAM 38/93;
B.~A.~Kniehl and A.~Sirlin, DESY-preprint DESY 93-194.

\bibitem{sv} B.~H.~Smith and M.~B.~Voloshin, UMN-TH-1241/94,
TPI-MINN-94/5-T.

\bibitem{qcd} K.~G.~Chetyrkin, Phys. Lett. B307(1993)169;
K.~G.~Chetyrkin and A.~Kwiatkowski, Phys. Lett. B305 (1993)285;
S.~A.~Larin, T.~van~Ritbergen and J.~A.~M.~Vermaseren,
Nikhef Preprint NIKHEF-H/93-26;
K.~G.~Chetyrkin and J.~H.~K\"uhn, Phys. Lett. B308 (1993)127.

\bibitem{priv} F.~Jegerlehner, private communication.
\end{thebibliography}
\end{document}